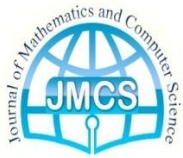 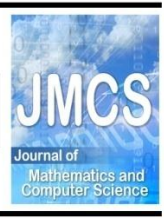

# Ergodicity of Fuzzy Markov Chains Based on Simulation Using Sequences

Behrouz Fathi Vajargah, Maryam Gharehdaghi
*Department of Statistics, University of Guilan, Rasht, Iran,*
*fathi@guilan.ac.ir*
*gharehdaghi.m@gmail.com*

*Abstract*

As shown in [1], we reduce periodicity of fuzzy Markov chains using the Halton quasi-random generator. In this paper, we employ two different quasi-random sequences namely Faure and Kronecker to generate the membership values of fuzzy Markov chain. Using simulation it is revealed that the number of ergodic fuzzy Markov chain simulated by Kronecker sequences is more than the one obtained by Faure sequences.

**Keywords:** Ergodic Fuzzy Markov Chains, Faure Sequence, Kronecker Sequence, Quasi-Random Sequences.

## 1. Introduction

Random simulation has long been a very popular and well studied field of mathematics. There exists a wide range of applications in biology, finance, insurance, physics and many others. So simulations of random numbers are crucial. In this paper, we describe two random number algorithms namely Faure and Kronecker [2]. Also, we define fuzzy Markov chains based on possibilities. Fuzzy Markov chains approaches are given by Avrachenkov and Sanchez in [5]. We simulate fuzzy Markov chains using two quasi-random sequences algorithms and observe efficiency of them in ergodicity of fuzzy Markov chains. Kronecker sequences give us better results than Faure sequences.

## 2. Fuzzy Markov chains

Let $S = \{1, 2, \ldots, n\}$. A finite fuzzy set on $S$ is defined by a mapping $x$ from $S$ to $[0,1]$ represented by a vector $x = \{x_1, x_2, \ldots, x_n\}$, with $0 \leq x_i \leq 1$, $i \in S$. Here, $x_i$ is the membership function that a state $i$ has regarding a fuzzy set $S$, $i \in S$. A fuzzy transition possibility matrix $P$ is defined in a metric space $S \times S$ by a matrix $\{\mu_{ij}\}_{i,j=1}^m$ with $0 \leq \mu_{ij} \leq 1$, $i, j \in S$. $\mu_{ij}$ is the membership value.

We note that it dose not need elements of each row of the matrix $P$ to sum up to one. This fuzzy matrix $P$ allows to define all relations among the $m$ states of the fuzzy Markov chain at each time instant $t$, as follows [1, 5].

At each instant $t$, $t = 1, 2, \ldots, n$, the state of system is described by the fuzzy set $x^{(t)}$. The transition law of a fuzzy Markov chain is given by the fuzzy relational matrix $P$ at instant $t$, $t = 1, 2, \ldots, n$, as follows:

$$x_j^{(t+1)} = max_{i \in S}\{min\{x_j^{(t)}, \mu_{ij}\}\} \quad , j \in S, \tag{1}$$

$$x^{(t+1)} = x^{(t)} \circ P,$$

where $i$ and $j$, $i, j = 1, 2, \ldots, n$, are the initial and final states of the transition and $x^{(0)}$ is the initial distribution. Also,

$$\mu_{ij}^t = max_{k \in S}\{min\{\mu_{ik}, \mu_{kj}^{t-1}\}\} i, j \in S, \tag{2}$$

$$P^t = P \circ P^{t-1}.$$

Thomason in [7] shows that the powers of a fuzzy matrix are stable over the *max-min* operator. More information about powers of a fuzzy matrix can be found in [6,7]. Now, a stationary distribution of a fuzzy matrix is defined as follows.

**Definition 2.1**. Let the powers of the fuzzy transition matrix $P$ converge in $\tau$ steps to a non-periodic solution, then the associated fuzzy Markov chain is called aperiodic fuzzy Markov chain and $P^* = P^\tau$ is its stationary fuzzy transition matrix.

**Definition 2.2**. A fuzzy Markov chain is called strong ergodic if it is aperiodic and its stationary transition matrix has identical rows.

A fuzzy Markov chain is called weakly ergodic if it is aperiodic and its stationary transition matrix is stable with no identical rows.

## 3. Quasi-Random Sequences for generating $\mu$

In this section, we present quasi-random number generation. By random numbers, we mean random varieties of the uniform *U(0, 1)* distribution. More complex distributions can be generated with uniform varieties and rejection or inversion methods. Quasi-random number generation aims to be deterministic. An easy example of quasi-random points is these sequence of $n$ terms given by $(\frac{1}{2n}, \frac{3}{2n}, \ldots, \frac{2n-1}{2n})$.

This sequence has a discrepancy $\frac{1}{n}$, see Niederreiter [2] for details. The problem with this finite sequence is it depends on $n$. If we want different point numbers, we need to recompute the whole sequence. In the following, we will on work the first $n$ points of an in finite sequence in order to use previous computation if we increase $n$. Moreover we introduce the notion of discrepancy on a finite sequence $(u_i)_{1 \leq i \leq n}$. In the above example, we are able to calculate exactly the discrepancy. With in finite sequence, this is no longer possible. Thus, we will only try to estimate asymptotic equivalents of discrepancy. The discrepancy of the average sequence of points is governed by the law of the iterated logarithm:

$$\lim sup_{n \to +\infty} \frac{\sqrt{n} D_n}{\sqrt{\log \log n}} = 1, \qquad (3)$$

which leads to the following asymptotic equivalent for $D_n$ [2,3]

$$D_n = O\left(\sqrt{\frac{\log \log n}{n}}\right). \qquad (4)$$

### 3.1  Faure Sequences

The Faure sequences is also based on the decomposition of integers into prime-basis but they have two differences: it uses only one prime number for basis and it permutes vector elements from one dimension to another. The basis prime number is chosen as the smallest prime number greater than the dimension d,i.e. 3 when $d = 2$, 5 when $d= 3$ or 4 etc. We decompose integer $n$ into the $p$-basis:

$$n = \sum_{j=1}^{k} a_j p^j. \qquad (5)$$

Let $a_{1,j}$ be integer $a_j$ used for the decomposition of $n$. Now we define a recursive permutation of

$a_j$:

$$\forall 2 \leq D \leq d, a_{D,j} = \sum_{j=1}^{k} C_j^i a_{D-1,j} \mod p, \quad (6)$$

Where $C_j^i$ denotes standard combination $j!/i!\,(j-i)!$. Then we take the radical-inversion $\phi_p(a_{D,1}, \ldots, a_{D,k})$ defined as

$$\phi(a_1, \ldots, a_k) = \sum_{j=1}^{k} \frac{a_j}{p^{j+1}}, \quad (7)$$

which is the same as above for $n$ defined by $a_{D,i}$'s. Finally the ($d$-dimensional) Faure sequence is defined by

$$(\phi_p(a_{1,1}, \ldots, a_{1,k}), \ldots, \phi_p(a_{d,1}, \ldots, a_{d,k})) \in I^D. \quad (8)$$

In the d-dimensional case, we work in 3-basis, first terms of the sequence are listed in table 1[2].

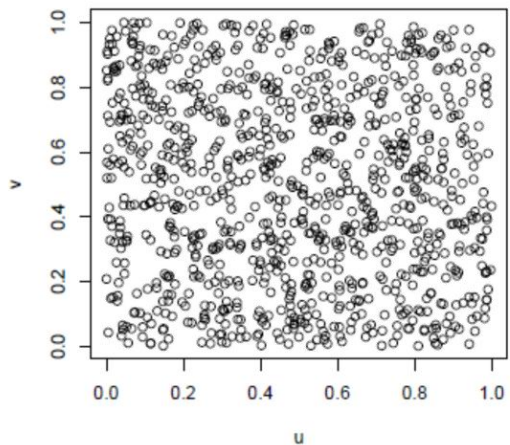

**Figure1. Faure Sequences**

**Table 1. Faure first terms**

| $n$ | $a_{13}a_{12}a_{11}^{2}$ | $a_{23}a_{22}a_{21}$ | $\phi(a_{13}..)$ | $\phi(a_{23}..)$ |
|---|---|---|---|---|
| 0 | 000 | 000 | 0 | 0 |
| 1 | 001 | 001 | 1/3 | 1/3 |
| 2 | 002 | 002 | 2/3 | 2/3 |
| 3 | 010 | 012 | 1/9 | 7/9 |
| 4 | 011 | 010 | 4/9 | 1/9 |
| 5 | 012 | 011 | 7/9 | 4/9 |
| 6 | 020 | 021 | 2/9 | 5/9 |
| 7 | 021 | 022 | 5/9 | 8/9 |
| 8 | 022 | 020 | 8/9 | 2/9 |
| 9 | 100 | 100 | 1/27 | 1/27 |
| 10 | 101 | 101 | 10/27 | 10/27 |
| 11 | 102 | 102 | 19/27 | 19/27 |
| 12 | 110 | 112 | 4/27 | 22/27 |
| 13 | 111 | 110 | 12/27 | 4/27 |
| 14 | 112 | 111 | 22/27 | 12/27 |

### 3.2 Kronecker Sequences

Another kind of quasi-random sequence uses irrational number and fractional part. The fractional part of a real $x$ is denoted by $\{x\} = x - [x]$. The infinite sequence $(n\{\alpha\})_{n\leq 0}$ has abound for its discrepancy

$$D_n \leq C \frac{1 + \log n}{n}. \qquad (9)$$

This family of infinite sequence $(n\{\alpha\})_{n\leq 0}$ is called the Kronecker sequence. A special case of the Kronecker sequence is the Torus algorithm where irrational number $\alpha$ is a square root of a prime number. The *n*th term of the *d*-dimensional Torus algorithm is defined by

$$\left(n\{\sqrt{p_1}\}, n\{\sqrt{p_2}\}, \dots, n\{\sqrt{p_d}\}\right) \in I^D, \qquad (10)$$

where $(p_1, p_2, \dots, p_d)$ are prime numbers, generally the first *d* prime numbers. With the previous inequality, we can derive an estimate of the Torus algorithm discrepancy

$$O\left(\frac{1 + \log n}{n}\right). \qquad (11)$$

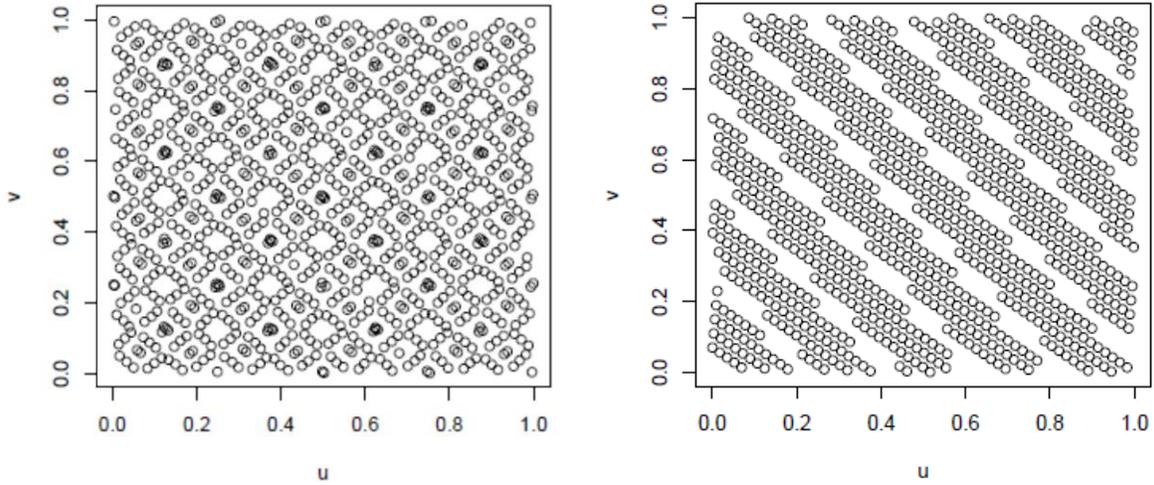

**Figure 2. Kronecker and Torus Sequences**

## 4. Simulation

We present a simulation on fuzzy Markov chains to identify some characteristics about their behavior based on matrix analysis. The Size of fuzzy Markov chains denoted by $n = \{5, 50, 100, 1000\}$.

All the entries $\mu_{ij}$ of the fuzzy transition matrix of fuzzy Markov chains are obtained by using the Faure sequences and Kronecker sequences.

The Table 2 and 3 show that the number of ergodic fuzzy Markov chain simulated by Kronecker sequences is more than the one obtained by Faure sequences.

**Table 2. Number of ergodic fuzzy Markov chains using Faure sequences**

| Size | Number of Ergodic fuzzy Markov chains |
|---|---|
| $n=5$ | 3 |
| $n=50$ | 35 |
| $n=100$ | 75 |
| $n=1000$ | 831 |

Table 3. Number of ergodic fuzzy Markov chains
using Kronecker sequences

| Size | Number of Ergodic fuzzy Markov chains |
|---|---|
| *n=5* | 4 |
| *n=50* | 41 |
| *n=100* | 88 |
| *n=1000* | 903 |

## 6. Conclusion

This work is an application of quasi-random sequences in investigating of ergodicity of fuzzy Markov chains. We compare two different quasi-random sequences. In term of ergodicity, Kronecker sequences give us better results than Faure sequences. Therefore, we can employ our approach to generate ergodic fuzzy Markov chains. Now need to say that such fuzzy Markov chains have wide applications in other fields such as image processing. We hope to employ the described approach in the near future.

## References

[1] B. Fathi, M. Gharehdaghi, Reducing periodicity of fuzzy Markov chains based on simulation using Halton sequences, accepted (2013).

[2] H. Niederreiter, Quasi-Monte Carlo methods and pseudo-random numbers, Bulletin of the American Mathematical Society 84(6) 1978.

[3] J.E. Gentle, Random Number Generation and Monte Carlo Methods. Springer, 2005

[4] J.J. Buckley, Fuzzy Probability and Statistics, F. Verlag, Ed. Springer-Verlag, 2004.

[5] K.E. Avrachenkov, E. Sanchez, Fuzzy Markov chains, IPMU. Spain (2000) 1851-1856.

[6] M. Gavalec, Computing orbit period in max-min algebra, Discrete Applied Mathematics 100 (2000) 49-65.

[7] M. Thomason, Convergence of powers of a fuzzy matrix, Journal of mathematical analysis and applications 57 (1977) 476-480.